\documentclass[12pt,thmsa]{article}
\usepackage{amssymb}

\input{tcilatex}
\begin{document}

\author{Harald Grosse$^a$, Karl-Georg Schlesinger$^b$ \\
$^a$Institute for Theoretical Physics\\
University of Vienna\\
Boltzmanngasse 5\\
A-1090 Vienna, Austria\\
e-mail: grosse@doppler.thp.univie.ac.at\\
$^b$Institute for Theoretical Physics\\
University of Vienna\\
Boltzmanngasse 5\\
A-1090 Vienna, Austria\\
e-mail: kgschles@esi.ac.at}
\title{On deformation theory of quantum vertex algebras}
\date{}
\maketitle

\begin{abstract}
We study an algebraic deformation problem which captures the data of the
general deformation problem for a quantum vertex algebra. We derive a system
of coupled equations which is the counterpart of the Maurer-Cartan equation
on the usual Hochschild complex of an associative algebra. We show that this
system of equations results from an action principle. This might be the
starting point for a perturbative treatment of the deformation problem of
quantum vertex algebras. Our action generalizes the action of the
Kodaira-Spencer theory of gravity and might therefore also be of relevance
for applications in string theory.
\end{abstract}

\section{Introduction}

In \cite{Bor} Borcherds has developed an abstract notion of vertex algebra
as a certain singular commutative ring in a symmetric tensor category. He
introduces a quantum vertex algebra as a singular braided commutative ring
in a symmetric tensor category. He also remarks that, in addition, one could
allow for the tensor category to be no longer symmetric but also braided.
This is supposed to be of relevance for vertex operators with non-integer
exponents in the formal series and should therefore be important for string
theory with background fields (see \cite{KO}). We will even drop the
requirement of braidings and allow the inner ring, as well, as the tensor
product of the outer category to become arbitrarily noncommutative. A
generalization of vertex algebras where the inner ring has no longer to be
braided has been developed under the name of field algebras in \cite{BK}. We
will drop most of the structure of a quantum vertex algebra and just
consider the essential ingredients of the deformation problem, i.e we keep
just those structures which can be deformed. In consequence, we consider a $%
\Bbb{C}$-linear, additive, monoidal category $\left( \mathcal{C},\bullet
,\otimes \right) $ with composition $\bullet $ and tensor product $\otimes $
together with an inner associative ring $R$. Finally, we make two additional
generalizations:\ 

\bigskip

\begin{itemize}
\item  In addition to $\otimes $ and the product on $R$, we also allow the
composition $\bullet $ of $\mathcal{C}$ to become deformed.

\item  We pass from an inner associative ring $R$ to an inner bialgebra
object $B$ in $\mathcal{C}$ with an associative product and a coassociative
coproduct. In \cite{BD} chiral algebras are introduced as a coordinate
independent generalization of vertex algebras. In the axiomatization used
there, chiral algebras appear as a generalization of universal envelopes of
Lie algebras in a certain monoidal category. One could consider in this case
deformations of the outer category or of the inner algebra where - in the
spirit of the theory of quantum groups - one could think of passing from the
deformation theory of Lie algebras to the more general deformation theory of
Hopf algebras. By passing from $R$ to $B$ as the inner object, we want to
include the essential data of the deformation problem of this case, too.
\end{itemize}

\bigskip

We can therefore think of the deformation problem to be studied, here, as
the most general deformation problem for the data of a quantum vertex
algebra (we will keep speaking of a quantum vertex algebra even in the most
general case where an inner bialgebra object is allowed and $\otimes $ is
noncommutative). We will state a conjecture below that even generalizations
of this deformation problem to higher category theory should not be
possible. As an additional requirement on the structure of $\mathcal{C}$, we
will assume that there exists a forgetful functor to the category of vector
spaces over $\Bbb{C}$ (where we allow vector spaces to become infinite
dimensional).

\bigskip

For an associative algebra $A$ over $\Bbb{C}$, the deformations of the
product $m$ to a new associative product are given by the solutions of the
Maurer-Cartan equation on the Hochschild complex. More precisely, let $%
\alpha $ be a bilinear map on $A$ and $m+\alpha $ the deformed associative
product. The Hochschild complex of $A$ consists of all linear maps 
\[
\Phi :A^{\otimes ^n}\rightarrow A 
\]
Call $n$ the degree of $\Phi $. On the Hochschild complex there is a product 
$\circ $ given for $\Phi _1$ of degree $n_1$ and $\Phi _2$ of degree $n_2$
by 
\begin{eqnarray*}
&&\Phi _1\circ \Phi _2\left( a_1,...,a_{n_1+n_2-1}\right) \\
&=&\sum_{k=0}^{n_1-1}\left( -1\right) ^{k\left( n_2-1\right) }\Phi _1\left(
a_1,...,a_k,\Phi _2\left( a_{k+1},...,a_{k+n_2}\right)
,a_{k+n_2+1},...,a_{n_1+n_2-1}\right)
\end{eqnarray*}
i.e. $\Phi _1\circ \Phi _2$ is of degree $n_1+n_2-1$. The Gerstenhaber
bracket $\left[ ,\right] $ on the Hochschild complex is the graded
commutator of $\circ $, i.e. 
\[
\left[ \Phi _1,\Phi _2\right] =\Phi _1\circ \Phi _2-\left( -1\right)
^{\left( n_1-1\right) \left( n_2-1\right) }\Phi _1\circ \Phi _2 
\]
Observe that $\left[ ,\right] $ lowers the degree by one, i.e. it is what is
commonly called an antibracket. We can now rewrite associativity of $m$ as 
\[
\left[ m,m\right] =0 
\]
because 
\begin{eqnarray*}
\left[ m,m\right] \left( a_1,a_2,a_3\right) &=&2\ \left( m\circ m\right)
\left( a_1,a_2,a_3\right) \\
&=&m\left( m\left( a_1,a_2\right) ,a_3\right) -m\left( a_1,m\left(
a_2,a_3\right) \right)
\end{eqnarray*}
Now, the requirement that $m+\alpha $ should be associative, again, can be
stated as 
\begin{eqnarray*}
0 &=&\left[ m+\alpha ,m+\alpha \right] \\
&=&\left[ m,m\right] +2\ \left[ m,\alpha \right] +\left[ \alpha ,\alpha
\right]
\end{eqnarray*}
Using that $\left[ m,m\right] =0$, this gives 
\[
\left[ m,\alpha \right] +\frac 12\left[ \alpha ,\alpha \right] =0 
\]
Defining 
\[
d_m\Phi =\left[ m,\Phi \right] 
\]
one checks that $d_m$ defines a differential on the Hochschild complex, i.e. 
$\left( d_m\right) ^2=0$. The cohomology with respect to $d_m$ is the
Hochschild cohomology of $A$.

So, the requirement that $\alpha $ gives a deformation of $m$ which
preserves associativity can be stated as 
\[
d_m\alpha +\frac 12\left[ \alpha ,\alpha \right] =0 
\]
which is the \textit{Maurer-Cartan equation} on the Hochschild complex of $A$%
. We will show below that in the general deformation problem for a quantum
vertex algebra, the Maurer-Cartan equation is generalized to a coupled
system of differential equations.

The Maurer-Cartan equation is the starting point for a perturbative solution
of the deformation problem. Let 
\begin{equation}
\alpha =\sum_{n=1}^\infty t^n\alpha _n  \label{1}
\end{equation}
where the $\alpha _n$ are bilinear operators on $A$, again, and $t$ is the
deformation parameter, i.e. in the perturbative treatment we consider $%
\alpha $ as a formal power series in $t$ and the deformation of $\left(
A,m\right) $ as a deformation over $\Bbb{C}\left[ \left[ t\right] \right] $.
The first order deformations $\alpha _1$ are parametrized by second
Hochschild cohomology $HH^2\left( A\right) $ of $A$. Inserting (\ref{1})
into the Maurer-Cartan equation, we get 
\begin{equation}
d_m\alpha _n=-\frac 12\sum_{k+l=n}\left[ \alpha _k,\alpha _l\right]
\label{2}
\end{equation}
for all $n$. Since $n\geq 1$, we get $k,l\leq n-1$ on the right hand side
and we can use (\ref{2}) to solve the Maurer-Cartan equation perturbatively
(in principle; the actual construction of the solution turns out to be a
highly involved problem, see \cite{Kon 1997} for a solution in the special
case where $A$ is the algebra of smooth functions on a manifold).

Now, let $M$ be a complex algebraic variety and suppose that $M$ is a
Calabi-Yau variety. Consider the Hochschild complex of the holomorphic
structure sheaf $\mathcal{O}_M$ of $M$. This has a similar structure as the
Hochschild complex of an associative algebra. The Hochschild cohomology $%
HH^n\left( \mathcal{O}_M\right) $of $\mathcal{O}_M$ decomposes as 
\begin{equation}
HH^n\left( \mathcal{O}_M\right) \cong \oplus _{p+q=n}H^p\left(
\bigwedge^qT_M\right)  \label{3}
\end{equation}
where $H^p$ denotes usual sheaf cohomology and $T_M$ is the holomorphic
tangent bundle of $M$. So, the Hochschild cohomology of $\mathcal{O}_M$ is
given as the cohomology of the exterior powers of the tangent bundle. The
first order deformations of the pointwise product structure of $\mathcal{O}%
_M $ are given by the component $HH^2\left( \mathcal{O}_M\right) $. This
includes, especially, the component with $p=q=1$ which parametrizes the
first order Kodaira-Spencer deformations of the complex structure of $M$.
The finite (i.e. noninfinitesimal) Kodaira-Spencer deformations are given by
the solutions of the Kodaira-Spencer equation 
\[
\overline{\partial }\gamma +\frac 12\left[ \gamma ,\gamma \right] =0 
\]
where, in accordance with (\ref{3}), $\gamma $ is a $\left( 0,1\right) $%
-form valued vector field and $\left[ ,\right] $ denotes the commutator of
vector fields and wedging. Observe that the Kodaira-Spencer equation is
formally a Maurer-Cartan equation, again. In \cite{BCOV} it was shown that
the Kodaira-Spencer equation can be derived from an action principle and
that the tree level expansion of the resulting field theory (\textit{%
Kodaira-Spencer theory of gravity}) precisely solves the Kodaira-Spencer
equation in a treatment completely analogous to (\ref{2}). The
BV-quantization of the theory then shows that one has to include
nonclassical ghost numbers which mathematically results in passing from the
Kodaira-Spencer deformations to the total Hochschild complex of $\mathcal{O}%
_M$ with the cohomology given by (\ref{3}) (in physics terminology passing
from the purely marginal deformations given by Kodaira-Spencer theory to the
extended moduli space of \cite{Wit}, see \cite{Kon 1994} for a mathematical
approach to this space). The distinction between the complex and the
cohomology on the mathematical side corresponds to the distinction between
off-shell and on-shell states in physics (see e.g. \cite{HM} for a detailed
exposition of this point). The Maurer-Cartan equation on the Hochschild
complex corresponds to the Master-equation of BV-quantization, the
Gerstenhaber bracket corresponds to the BV-antibracket, and the BV-action
functional reduces to the classical action if one considers only classical
fields because the Maurer-Cartan equation on (\ref{3}) reduces to the
Kodaira-Spencer equation if one considers only Kodaira-Spencer deformations.
So, the essential requirements of BV-quantization (Master-equation and the
correct reduction to the classical action on classical fields) are fulfilled
just by the fact that the Kodaira-Spencer deformation can be embedded into
the larger deformation theory on the Hochschild complex of $\mathcal{O}_M$
(with Maurer-Cartan type equations being preserved).

We show below that the coupled system of differential equations which we
derive for the general deformation problem of quantum vertex algebras can
likewise be derived from an action principle. This action can be understood
as generalizing the action of the Kodaira-Spencer theory of gravity. We
formulate a list of goals which would be interesting to reach in a
perturbative (tree level, as well, as loop expansion) treatment of the
resulting field theory. At present the perturbative study of the theory is
limited by the fact that one would first have to find a suitable gauge
fixing for the action (corresponding to the Tian gauge in the
Kodaira-Spencer theory of gravity).

\bigskip

The paper is structured as follows: In section 2, we consider the
deformation problem for the outer category $\left( \mathcal{C},\bullet
,\otimes \right) $, i.e. $\bullet $ and $\otimes $ receive deformations. In
section 3, we include the inner bialgebra object $B$ into the consideration.
We will see that one needs an additional assumption - which basically says
that a reconstruction theorem between $B$ and its category of
representations in $\mathcal{C}$ is supposed to hold - in order to be able
to treat this case as an algebraic deformation problem. Under this
assumption, the deformation problem can be reformulated as that for two
monoidal categories $\left( \mathcal{M},\bullet ,\otimes \right) $ and $%
\left( \mathcal{C},\bullet ,\otimes \right) $ with a monoidal forgetful
functor 
\[
\mathcal{F}:\mathcal{M}\rightarrow \mathcal{C} 
\]
where both compositions $\bullet $ and both tensor products $\otimes $ and
the monoidal functor $\mathcal{F}$ become deformed. Here, and in the sequel,
we will use the same symbol for the compositions in $\mathcal{M}$ and $%
\mathcal{C}$ (and likewise for the tensor products) if no confusion is
possible. In section 4, we formulate the list of goals mentioned above.
Section 5 contains some concluding remarks.

\bigskip

\textit{A disclaimer:} We certainly do not claim that the structures which
we study in this paper constitute a notion of quantum vertex algebras. Much
more structure is needed for this. For part of this structure, one would
expect the existence of theorems saying that the deformation of these
structural components canonically goes along with the deformations
considered, here (much the same way as e.g. the counit in a Hopf algebra
canonically goes along with the deformations of the product and coproduct).
But the proof of such theorems is a nontrivial matter which needs further
investigation. When we speak of the deformation problem of quantum vertex
algebras in our context, we only want to point to the following fact: Vertex
algebras can be considered as singular commutative rings in certain monoidal
categories. In this sense, the deformation problem which we consider is the
most general deformation problem for those part of structure which are the
essential ones for the deformation of vertex algebras. Given the highly
nontrivial nature of the question of deformations of quantum vertex
algebras, we believe that such a boiled down approach is justified.

\bigskip

\section{Deformation of the monoidal category}

Let 
\[
\mathcal{A}:=Morph\left( \mathcal{C}\right) 
\]
denote the class of morphisms of $\mathcal{C}$. Let $\boxtimes $ denote the
tensor product of $\Bbb{C}$-linear tensor categories as used e.g. in \cite
{CF}, i.e. we can understand a bilinear functor on $\mathcal{C}$ as a linear
functor from $\mathcal{C}\boxtimes \mathcal{C}$ to $\mathcal{C}$. We also
denote by $\boxtimes $ the corresponding tensor product on $\mathcal{A}$.
Then let $\mathcal{TA}$ denote the categorical tensor algebra of multilinear
maps on $\mathcal{A}$, i.e. 
\[
\mathcal{TA}=\oplus _{n=0}^\infty \left\{ \Phi :\boxtimes ^n\mathcal{A}%
\rightarrow \mathcal{A}\ linear\right\} 
\]
where the definition of the categorical direct sum is obvious.

In deforming the composition $\bullet $ to $\widetilde{\bullet }$ we will
assume that the domain and codomain maps on $\mathcal{C}$ remain undeformed,
i.e. $a\bullet b$ is defined iff $a\widetilde{\bullet }b$ is defined. For
the deformation of $\otimes $ to $\widetilde{\otimes }$, we make the
following assumption: We assume that this deformation does not change the
tensor product of objects on a $\Bbb{C}$-linear level, i.e the object maps
of $\otimes $ and $\widetilde{\otimes }$ should agree after applying the
monoidal forgetful functors to the category $\mathcal{V}$ of (finite and
infinite dimensional) vector spaces to both (remember that we suppose such
forgetful functors always to exist, see the introduction). Especially, this
means that after applying forgetful functors the domains and codomains of $%
a\otimes b$ and $a\widetilde{\otimes }b$ agree. In addition, we assume that
we can fully study the deformation problem for $\otimes $ by studying the
deformation problem for the morphism map of $\otimes $.

We write 
\[
\widetilde{\bullet }=\bullet +f 
\]
and 
\[
\widetilde{\otimes }=\otimes +g 
\]
where $f$ and $g$ are linear 
\[
f,g:\mathcal{A}\boxtimes \mathcal{A}\rightarrow \mathcal{A} 
\]
and addition + can be understood as componentwise addition on the $Hom_{%
\mathcal{C}}\left( ,\right) $-sets after applying the forgetful functor to $%
\mathcal{V}$.

We start with the consideration of the deformation problem for $\bullet $.
The requirement is that $\widetilde{\bullet }$ should be associative, again.
For $a,b,c$ in $\mathcal{A}$ and composable in the required order, this
leads to 
\[
\left( a\widetilde{\bullet b}\right) \widetilde{\bullet }c=a\widetilde{%
\bullet }\left( b\widetilde{\bullet }c\right) 
\]
and therefore to 
\[
\left[ \bullet ,f\right] +\frac 12\left[ f,f\right] =\left[ \bullet
,f\right] +f\circ f=0 
\]
where $\circ $ and $\left[ ,\right] $ on $\mathcal{TA}$ are defined in
complete analogy to the case of the Hochschild complex of an associative
algebra (see the introduction). Observe, here, that $f\left( a,b\right) $ is
defined iff $a\bullet b$ is defined and that $\bullet $ is linear, 
\[
\bullet :\mathcal{A}\boxtimes \mathcal{A}\rightarrow \mathcal{A} 
\]
For $\Phi $ in $\mathcal{TA}$ define $d_{\bullet }$ by 
\[
d_{\bullet }\Phi =\left[ \bullet ,\Phi \right] 
\]
We have $\left( d_{\bullet }\right) ^2=0$ and $d_{\bullet }$ defines a
differential on $\mathcal{TA}$. Then associativity of $\widetilde{\bullet }$
can be formulated as 
\begin{equation}
d_{\bullet }f+f\circ f=0  \label{4}
\end{equation}
which is a Maurer-Cartan type equation.

Let us now pass to the deformation problem for $\otimes $. We assume that
all the monoidal categories which we consider are strict monoidal. This is
possible by a coherence theorem of Mac Lane (see \cite{Mac}). Besides making
calculations simpler, this assumption might also be justified on a deeper
level: By the very existence of the coherence theorem, one expects that when
deriving the deformation problem from a field theory action principle, the
more general case of a nontrivial associator should automatically arise by
including BRST-exact terms. Then the restriction imposed upon the
deformation problem is, again, associativity of $\widetilde{\otimes }$,
leading to 
\begin{equation}
d_{\otimes }g+g\circ g=0  \label{5}
\end{equation}
where $d_{\otimes }$ is defined by 
\[
d_{\otimes }\Phi =\left[ \otimes ,\Phi \right] 
\]
for $\Phi $ in $\mathcal{TA}$. Observe that by the restriction to the
consideration of the morphism map of $\otimes $, we can, again, view $%
\otimes $ as linear, 
\[
\otimes :\mathcal{A}\boxtimes \mathcal{A}\rightarrow \mathcal{A} 
\]
What remains to be considered is the compatibility constraint on $\widetilde{%
\bullet }$ and $\widetilde{\otimes }$ in order that they define together a
monoidal category, again. For $a_1,a_2,a_3,a_4$ in $\mathcal{A}$ such that 
\[
\left( a_1\otimes a_2\right) \bullet \left( a_3\otimes a_4\right) 
\]
exists this means that 
\[
\left( a_1\widetilde{\otimes }a_2\right) \widetilde{\bullet }\left( a_3%
\widetilde{\otimes }a_4\right) =\left( a_1\widetilde{\bullet }a_3\right) 
\widetilde{\otimes }\left( a_2\widetilde{\bullet }a_4\right) 
\]
i.e. 
\begin{eqnarray*}
&&\left( \bullet +f\right) \left( \left( \otimes +g\right) \left(
a_1,a_2\right) ,\left( \otimes +g\right) \left( a_3,a_4\right) \right) \\
&=&\left( \otimes +g\right) \left( \left( \bullet +f\right) \left(
a_1,a_3\right) ,\left( \bullet +f\right) \left( a_2,a_4\right) \right)
\end{eqnarray*}
which by using compatibility of $\bullet $ and $\otimes $ is equivalent to 
\begin{eqnarray}
&&g\left( a_1,a_2\right) \bullet \left( a_3\otimes a_4\right) +\left(
a_1\otimes a_2\right) \bullet g\left( a_3,a_4\right)  \label{6} \\
&&+g\left( a_1,a_2\right) \bullet g\left( a_3,a_4\right)  \nonumber \\
&&+f\left( a_1\otimes a_2,a_3\otimes a_4\right) +f\left( a_1\otimes
a_2,g\left( a_3,a_4\right) \right)  \nonumber \\
&&+f\left( g\left( a_1,a_2\right) ,a_3\otimes a_4\right) +f\left( g\left(
a_1,a_2\right) ,g\left( a_3,a_4\right) \right)  \nonumber \\
&&-f\left( a_1,a_3\right) \otimes \left( a_2\bullet a_4\right) -\left(
a_1\bullet a_3\right) \otimes f\left( a_2,a_4\right)  \nonumber \\
&&-f\left( a_1,a_3\right) \otimes f\left( a_2,a_4\right)  \nonumber \\
&&-g\left( a_1\bullet a_3,a_2\bullet a_4\right) -g\left( a_1\bullet
a_3,f\left( a_2,a_4\right) \right)  \nonumber \\
&&-g\left( f\left( a_1,a_3\right) ,a_2\bullet a_4\right) -g\left( f\left(
a_1,a_3\right) ,f\left( a_2,a_4\right) \right)  \nonumber \\
&=&0  \nonumber
\end{eqnarray}
We will now discuss the structural properties of the terms in (\ref{6}).
Observe that (\ref{6}) contains terms of first, second, and third order in
the fields $f,g$ (especially, this means that these terms appear, upon
expanding $f$ and $g$ according to (\ref{1}), for the first time in the
respective order in the resulting expansion of (\ref{6})). We start with
considering the first order terms: There are six first order terms, three in
each of the fields $f$ and $g$. Taking only the three first order terms in $%
g $, we have 
\begin{equation}
\left( a_1\otimes a_2\right) \bullet g\left( a_3,a_4\right) -g\left(
a_1\bullet a_3,a_2\bullet a_4\right) +g\left( a_1,a_2\right) \bullet \left(
a_3\otimes a_4\right)  \label{7}
\end{equation}
Remember that $g$ is linear, 
\[
g:\mathcal{A}\boxtimes \mathcal{A}\rightarrow \mathcal{A} 
\]
But - as a consequence of our assumption that domain and codomain maps are
preserved in the deformation - we can locally on the $Hom$-sets replace $%
\boxtimes $ by $\otimes $ after applying the forgetful functor to $\mathcal{V%
}$ (alternatively, we can view this as evaluating an expression $a\boxtimes
b $ to $\otimes \left( a\boxtimes b\right) =a\otimes b$). We can then simply
rewrite (\ref{7}) to 
\begin{eqnarray*}
&&\left( a_1\otimes a_2\right) \bullet g\left( a_3\otimes a_4\right)
-g\left( \left( a_1\bullet a_3\right) \otimes \left( a_2\bullet a_4\right)
\right) +g\left( a_1\otimes a_2\right) \bullet \left( a_3\otimes a_4\right)
\\
&=&\left( a_1\otimes a_2\right) \bullet g\left( a_3\otimes a_4\right)
-g\left( \left( a_1\otimes a_2\right) \bullet \left( a_3\otimes a_4\right)
\right) +g\left( a_1\otimes a_2\right) \bullet \left( a_3\otimes a_4\right)
\end{eqnarray*}
where we now interpret $g$ as a function of one variable on the tensor
product. Defining 
\begin{eqnarray*}
&&d_{\bullet }^{\otimes ^2}g\left( a_1,a_2,a_3,a_4\right) \\
&=&\left( a_1\otimes a_2\right) \bullet g\left( a_3\otimes a_4\right)
-g\left( \left( a_1\otimes a_2\right) \bullet \left( a_3\otimes a_4\right)
\right) +g\left( a_1\otimes a_2\right) \bullet \left( a_3\otimes a_4\right)
\end{eqnarray*}
and similarly for a general $n$-variable function $\Phi $ on the tensor
product, we see that $d_{\bullet }^{\otimes ^2}$is just the Hochschild
differential $d_{\bullet }$ with respect to $\bullet $, as introduced above,
lifted to the twofold tensor product (hence the notation $d_{\bullet
}^{\otimes ^2}$). It therefore immediately follows that 
\[
\left( d_{\bullet }^{\otimes ^2}\right) ^2=0 
\]
Hence, the three terms in (\ref{7}) can be written as 
\[
d_{\bullet }^{\otimes ^2}g\left( a_1,a_2,a_3,a_4\right) 
\]
The three first order terms in $f$ are 
\begin{equation}
-\left( a_1\bullet a_3\right) \otimes f\left( a_2,a_4\right) +f\left(
a_1\otimes a_2,a_3\otimes a_4\right) -f\left( a_1,a_3\right) \otimes \left(
a_2\bullet a_4\right)  \label{8}
\end{equation}
By interchanging the roles of $\bullet $ and $\otimes $, we can - in
complete analogy to the preceding argument - rewrite (\ref{8}). This time,
we imagine $a\boxtimes b$ as being evaluated to $\bullet \left( a\boxtimes
b\right) =a\bullet b$ in order to interpret $f$ as a function of a single
variable. With 
\begin{eqnarray*}
&&d_{\otimes }^{\otimes ^2}f\left( a_1,a_2,a_3,a_4\right) \\
&=&\left( a_1\bullet a_3\right) \otimes f\left( a_2\bullet a_4\right)
-f\left( \left( a_1\otimes a_2\right) \bullet \left( a_3\otimes a_4\right)
\right) +f\left( a_1\bullet a_3\right) \otimes \left( a_2\bullet a_4\right)
\\
&=&\left( a_1\bullet a_3\right) \otimes f\left( a_2\bullet a_4\right)
-f\left( \left( a_1\bullet a_3\right) \otimes \left( a_2\bullet a_4\right)
\right) +f\left( a_1\bullet a_3\right) \otimes \left( a_2\bullet a_4\right)
\end{eqnarray*}
we can rewrite the three first order terms in $f$ as 
\[
-d_{\otimes }^{\otimes ^2}f\left( a_1,a_2,a_3,a_4\right) 
\]
Observe that we keep the upper index $\otimes ^2$ for the differential
lifted to the two fold tensor product since this is what occurs for the
linear spaces before evaluating with $\bullet $. $d_{\otimes }^{\otimes ^2}$
is the Hochschild differential $d_{\otimes }$ from above, lifted to the two
fold tensor product. Especially, we have 
\[
\left( d_{\otimes }^{\otimes ^2}\right) ^2=0 
\]
Let us next turn to the consideration of the second order terms. There are,
again, six terms: There is one term each which is quadratic in $f$,
respectively $g$, and four terms which contain $f$ and $g$. From the mixed
terms, two have $g$ left of $f$ and two vice versa. Let us start with the
consideration of the latter ones: These are 
\[
f\left( g\left( a_1,a_2\right) ,a_3\otimes a_4\right) +f\left( a_1\otimes
a_2,g\left( a_3,a_4\right) \right) 
\]
Rewriting this, again as 
\[
f\left( g\left( a_1\otimes a_2\right) ,a_3\otimes a_4\right) +f\left(
a_1\otimes a_2,g\left( a_3\otimes a_4\right) \right) 
\]
we can view $g$ as a function of one variable and $f$ as a function of two
variables on the twofold tensorproduct. Using the composition $\circ $ on $%
\mathcal{TA}$, we can write this as 
\[
\left( f\circ ^2g\right) \left( a_1,a_2,a_3,a_4\right) 
\]
where the notation $\circ ^2$- which should in a more detailed form be
written as $\circ ^{\otimes ^2}$ - indicates the lift to the twofold
tensorproduct, again. Similarly, we can rewrite the other two mixed terms as 
\[
-\left( g\circ ^2f\right) \left( a_1,a_2,a_3,a_4\right) 
\]
So, the second order terms can altogether be written as 
\[
\left( \left( f\circ ^2g\right) -\left( g\circ ^2f\right) +g\bullet
g-f\otimes f\right) \left( a_1,a_2,a_3,a_4\right) 
\]

\bigskip

\begin{remark}
Here and in the sequel, we neglect any switch functions for the arguments as
they appear in terms like $f\otimes f$ where the arguments appear in the
order $a_1,a_3,a_2,a_4$ instead of $a_1,a_2,a_3,a_4$. These switches are
obvious from the requirement that composability in the undeformed category $%
\mathcal{C}$ has to be satisfied (and consequently also holds then in the
deformed setting) for all the expressions appearing.
\end{remark}

\bigskip

Finally, we turn to the consideration of the third order terms. In usual
Hochschild cohomology, one only has a differential (function of one
variable)\ and a bracket (function of two variables) on the complex. Here,
we encounter a new function of three variables which we introduce for 
\[
\Phi _1,\Phi _2,\Phi _3:\mathcal{A}^{\otimes ^2}\rightarrow \mathcal{A} 
\]
as 
\[
Comp\left( \Phi _1,\Phi _2,\Phi _3\right) \left( a_1,a_2,a_3,a_4\right)
=\Phi _1\left( \Phi _2\left( a_1,a_2\right) ,\Phi _3\left( a_3,a_4\right)
\right) 
\]
For general $\Phi _1,\Phi _2,\Phi _3$ of degrees $n_1,n_2,n_3$,
respectively, $Comp$ generalizes to an alternating sum of $\Phi _2$ and $%
\Phi _3$ inserted into $\Phi _1$ with different positions in the arguments
of $\Phi _1$, again. Using this three variable function $Comp$, we can
rewrite (\ref{6}) as 
\begin{eqnarray}
0 &=&d_{\otimes }^{\otimes ^2}f-d_{\bullet }^{\otimes ^2}g+f\otimes
f-g\bullet g-\left( f\circ ^2g\right) +\left( g\circ ^2f\right)  \label{9} \\
&&-Comp\left( f,g,g\right) +Comp\left( g,f,f\right)  \nonumber
\end{eqnarray}
So, the deformation problem of a monoidal category leads to the coupled
system of differential equations (\ref{4}), (\ref{5}), and (\ref{9}) as
replacing the Maurer-Cartan equation on the usual Hochschild complex.

\bigskip

\begin{remark}
By linearity in the three arguments, the third order terms involving $Comp$
are similar to the appearance of curvature terms. The only difference is
that the curvature tensor is antisymmetric with respect to the last two
indices while the two terms $Comp\left( f,g,g\right) $ and $Comp\left(
g,f,f\right) $ appearing are symmetric in the last two variables. Can one
understand the third order terms as true curvature contribution by taking
into account the grading on the deformation complex (i.e. a kind of
super-curvature)? If yes, does this imply that one no longer has a flat
connection as in \cite{Dub} on the moduli space of deformations? This would
be similar to the situation one has for the appearance of superpotentials in
string theory.
\end{remark}

\bigskip

Let us assume the existence of a trace on $\mathcal{TA}$ which we will write
as $\int $ . Then we claim that one can formally derive this coupled system
of differential equations from the action 
\begin{eqnarray}
S &\sim &\int \{f\circ d_{\bullet }f+\frac 23f\circ f\circ f+g\circ
d_{\otimes }g+\frac 23g\circ g\circ g  \label{10} \\
&&+\lambda \circ ^2[d_{\otimes }^{\otimes ^2}f-d_{\bullet }^{\otimes
^2}g+f\otimes f-g\bullet g-\left( f\circ ^2g\right) +\left( g\circ ^2f\right)
\nonumber \\
&&-Comp\left( f,g,g\right) +Comp\left( g,f,f\right) ]\}  \nonumber
\end{eqnarray}
with fields $f,g,\lambda $. Here, the field 
\[
\lambda :\mathcal{A}\otimes \mathcal{A}\rightarrow \mathcal{A} 
\]
enters as a Lagrange multiplier into (\ref{10}) in order to install the
constraint (\ref{9}).

\bigskip

\begin{remark}
For a bundle of quantum vertex algebras on a manifold $M$, we can view the
integral $\int $ as including the integration over $M$ besides the trace.
\end{remark}

\bigskip

Observe that variation of $S$ with respect to $f$ leads in $\mathcal{TA}$ to
an expression in degree three resulting from the first two terms and to an
expression in degree four for the rest. Therefore these two expressions have
to vanish separately and we get associativity of $\bullet +f$ from the first
two terms of $S$. Similarly, we get associativity of $\otimes +g$. The
action (\ref{10}) describes a system of two Chern-Simons like theories,
coupled by the constraint (\ref{9}).

\bigskip

\section{Inclusion of the inner bialgebra object}

We now want to generalize (\ref{10}) to include the deformation of the inner
bialgebra $B$ object, too. Suppose, first, that $B$ is only an inner algebra
object of $\mathcal{C}$ with associative product $\star $. Deforming $\star $
to $\widetilde{\star }$ gives another Maurer-Cartan type equation from the
requirement that $\widetilde{\star }$ should be associative, again. The
compatibility constraint with the deformation of $\mathcal{C}$ is that $%
\widetilde{\star }$ should be a morphism in the deformed monoidal category $%
\left( \widetilde{\mathcal{C}},\widetilde{\bullet },\widetilde{\otimes }%
\right) $. But the condition of being a morphism in $\widetilde{\mathcal{C}}$
is not a purely algebraic condition but one of the type saying that $%
\widetilde{\star }$ should behave well under all possible compositions with
morphisms of $\widetilde{\mathcal{C}}$. We therefore proceed in a slightly
different way which we are now going to explain. It turns out to be useful
to consider $B$ as a bialgebra object in $\mathcal{C}$, now, from the
beginning.

Let \textit{Rep}$_{\mathcal{C}}\left( B\right) $ be the category of
representations of $B$ on objects of $\mathcal{C}$. Since $B$ is a bialgebra
object in $\mathcal{C}$ and therefore carries a coproduct, \textit{Rep}$_{%
\mathcal{C}}\left( B\right) $ is a monoidal category. We will now make an
additional assumption: In the theory of usual Hopf algebras, there exist a
number of so called reconstruction theorems which state that under certain
conditions a Hopf algebra and its category of representations determine each
other up to isomorphism (see e.g. \cite{CP}). We will assume that in the
same way the inner bialgebra object $B$ can be reconstructed from its
category \textit{Rep}$_{\mathcal{C}}\left( B\right) $ of representations in $%
\mathcal{C}$. With this assumption, the deformation problem of the inner
bialgebra object $B$ turns into the deformation problem for the monoidal
category 
\[
\mathcal{M}=\mathit{Rep}_{\mathcal{C}}\left( B\right) 
\]
where deformations of the product on $B$ correspond to deformations of the
composition $\bullet $ in $\mathcal{M}$ while deformations of the coproduct
of $B$ correspond to deformations of the tensor product $\otimes $ of $%
\mathcal{M}$. Here and in the sequel we use the same symbols for composition 
$\bullet $ and tensor product $\otimes $ of $\mathcal{C}$ and $\mathcal{M}$
if no confusion is possible. Hence, we double the deformation structure
studied in the previous sections by studying the deformations of two
monoidal categories $\mathcal{C}$ and $\mathcal{M}$. As the compatibility
constraint - replacing the constraint that $\widetilde{\star }$ should be a
morphism of $\widetilde{\mathcal{C}}$ - we use the requirement that the
monoidal forgetful functor 
\[
\mathcal{F}:\mathcal{M}\rightarrow \mathcal{C} 
\]
should be deformed to a monoidal functor 
\[
\widetilde{\mathcal{F}}:\widetilde{\mathcal{M}}\rightarrow \widetilde{%
\mathcal{C}} 
\]
With 
\[
\widetilde{\bullet }=\bullet +f 
\]
and 
\[
\widetilde{\otimes }=\otimes +g 
\]
for the deformation of $\mathcal{C}$ as above and similarly 
\[
\widetilde{\bullet }=\bullet +\varphi 
\]
and 
\[
\widetilde{\otimes }=\otimes +\psi 
\]
for the deformation of $\mathcal{M}$, plus 
\[
\widetilde{\mathcal{F}}=\mathcal{F}+\Omega 
\]
for the deformation of $\mathcal{F}$, we have altogether five fields $%
f,g,\varphi ,\psi ,\Omega $. Here, $\Omega $ is viewed as a linear map
between the morphism classes of $\mathcal{M}$ and $\mathcal{C}$ since the
object classes remain undeformed after applying forgetful functors to $%
\mathcal{V}$ (linearity of $\Omega $ means, of course the local linearity on
the $Hom$-sets). We will now study the constraints arising for $\Omega $.
Since $\widetilde{\mathcal{F}}$ has to be a monoidal functor, again, we have 
\begin{equation}
\widetilde{\mathcal{F}}\left( a\widetilde{\bullet }b\right) =\widetilde{%
\mathcal{F}}\left( a\right) \widetilde{\bullet }\widetilde{\mathcal{F}}%
\left( b\right)  \label{11}
\end{equation}
and 
\begin{equation}
\widetilde{\mathcal{F}}\left( a\widetilde{\otimes }b\right) =\widetilde{%
\mathcal{F}}\left( a\right) \widetilde{\otimes }\widetilde{\mathcal{F}}%
\left( b\right)  \label{12}
\end{equation}
for the morphisms of $\widetilde{\mathcal{M}}$. Let us study (\ref{11}) in
detail. (\ref{11}) implies 
\begin{eqnarray}
0 &=&\mathcal{F}\left( \varphi \left( a,b\right) \right) +\Omega \left(
a\bullet b\right) -\mathcal{F}\left( a\right) \bullet \Omega \left( b\right)
\label{13} \\
&&-\Omega \left( a\right) \bullet \mathcal{F}\left( b\right) -f\left( 
\mathcal{F}\left( a\right) ,\mathcal{F}\left( b\right) \right)  \nonumber \\
&&+\Omega \left( \varphi \left( a,b\right) \right) -\Omega \left( a\right)
\bullet \Omega \left( b\right) -f\left( \mathcal{F}\left( a\right) ,\Omega
\left( b\right) \right) -f\left( \Omega \left( a\right) ,\mathcal{F}\left(
b\right) \right)  \nonumber \\
&&-f\left( \Omega \left( a\right) ,\Omega \left( b\right) \right)  \nonumber
\end{eqnarray}
Remember that, here, $\mathcal{F}$ is not a field but the constant
undeformed monoidal forgetful functor from $\mathcal{M}$ to $\mathcal{C}$. (%
\ref{13}) contains five terms of first order in the fields, four terms of
second order, and one term of third order. We start by considering the three
first order terms in $\Omega $. These are 
\[
\mathcal{F}\left( a\right) \bullet \Omega \left( b\right) -\Omega \left(
a\bullet b\right) +\Omega \left( a\right) \bullet \mathcal{F}\left( b\right) 
\]
Defining 
\begin{eqnarray*}
d_{\bullet }^{\mathcal{F}}\Phi \left( a_1,...,a_{n+1}\right) &=&\mathcal{F}%
\left( a_1\right) \bullet \Phi \left( a_2,...a_{n+1}\right) -\Phi \left(
a_1\bullet a_2,a_3,...,a_{n+1}\right) \\
&&\pm ...\pm \Phi \left( a_1,...,a_{n-1},a_n\bullet a_{n+1}\right) \pm \Phi
\left( a_1,...,a_n\right) \bullet \mathcal{F}\left( a_{n+1}\right)
\end{eqnarray*}
we have 
\[
d_{\bullet }^{\mathcal{F}}\Omega \left( a,b\right) =\mathcal{F}\left(
a\right) \bullet \Omega \left( b\right) -\Omega \left( a\bullet b\right)
+\Omega \left( a\right) \bullet \mathcal{F}\left( b\right) 
\]
One proves by calculation - using the functoriality of $\mathcal{F}$ - that 
\[
\left( d_{\bullet }^{\mathcal{F}}\right) ^2=0 
\]
We call $d_{\bullet }^{\mathcal{F}}$ the $\mathcal{F}$-twist of the
differential $d_{\bullet }$.

We write $\mathcal{F}\left( \varphi \right) \left( a,b\right) $ for $%
\mathcal{F}\left( \varphi \left( a,b\right) \right) $ since we just apply $%
\mathcal{F}$ after $\varphi $, here. Similarly, we write $f\left( \mathcal{F}%
^{\otimes ^2}\right) \left( a,b\right) $ for $f\left( \mathcal{F}\left(
a\right) ,\mathcal{F}\left( b\right) \right) $, indicating in the notation
that $\mathcal{F}$ is first applied in both components. Then the first order
terms of (\ref{13}) can be summarized to 
\begin{equation}
\mathcal{F}\left( \varphi \right) -f\left( \mathcal{F}^{\otimes ^2}\right)
-d_{\bullet }^{\mathcal{F}}\Omega  \label{14}
\end{equation}
Let us next consider the two second order terms in $f$ and $\Omega $. We
have 
\[
f\left( \Omega \left( a\right) ,\mathcal{F}\left( b\right) \right) +f\left( 
\mathcal{F}\left( a\right) ,\Omega \left( b\right) \right) 
\]
Compare this with the composition of $f$ of degree two and $\Omega $ of
degree one to 
\[
\left( f\circ \Omega \right) \left( a,b\right) =f\left( \Omega \left(
a\right) ,b\right) +f\left( a,\Omega \left( b\right) \right) 
\]
In complete analogy to the $\mathcal{F}$-twisted differential $d_{\bullet }^{%
\mathcal{F}}$, we can introduce the $\mathcal{F}$-twisted composition $\circ
^{\mathcal{F}}$and get 
\[
\left( f\circ ^{\mathcal{F}}\Omega \right) \left( a,b\right) =f\left( \Omega
\left( a\right) ,\mathcal{F}\left( b\right) \right) +f\left( \mathcal{F}%
\left( a\right) ,\Omega \left( b\right) \right) 
\]

\bigskip

\begin{remark}
Observe that the positions in which $\mathcal{F}$ appears in the $\mathcal{F}
$-twisted forms are determined by the requirement that the needed
compositions are well defined.
\end{remark}

\bigskip

Introducing $\Omega \left( \varphi \right) $ in complete analogy to the
notation for the first order terms, we can summarize the four second order
terms of (\ref{13}) to 
\begin{equation}
\Omega \left( \varphi \right) -\Omega \bullet \Omega -f\circ ^{\mathcal{F}%
}\Omega  \label{15}
\end{equation}
Using, again, the notation introduced for the first order terms, we can
rewrite the single third order term as $f\left( \Omega ^{\otimes ^2}\right) $%
. Together with (\ref{14}) and (\ref{15}), this leads to an equivalent
reformulation of (\ref{13}) as 
\begin{eqnarray}
0 &=&d_{\bullet }^{\mathcal{F}}\Omega +f\left( \mathcal{F}^{\otimes
^2}\right) -\mathcal{F}\left( \varphi \right) +f\circ ^{\mathcal{F}}\Omega
\label{16} \\
&&+\Omega \bullet \Omega -\Omega \left( \varphi \right) +f\left( \Omega
^{\otimes ^2}\right)  \nonumber
\end{eqnarray}
In a completely analogous way, we can treat the constraint (\ref{12}),
leading to 
\begin{eqnarray}
0 &=&d_{\otimes }^{\mathcal{F}}\Omega +g\left( \mathcal{F}^{\otimes
^2}\right) -\mathcal{F}\left( \psi \right) +g\circ ^{\mathcal{F}}\Omega
\label{17} \\
&&+\Omega \otimes \Omega -\Omega \left( \psi \right) +g\left( \Omega
^{\otimes ^2}\right)  \nonumber
\end{eqnarray}
In summary, for the complete deformation problem of the two monoidal
categories $\mathcal{C}$ and $\mathcal{M}$ together with the monoidal
functor 
\[
\mathcal{F}:\mathcal{M}\rightarrow \mathcal{C} 
\]
we get a doubled set of the three coupled differential equations of the
previous section plus the two constraints (\ref{16}) and (\ref{17}). So, in
total, we get a coupled system of eight differential equations which replace
for the general deformation problem of quantum vertex algebras the
Maurer-Cartan equation of the usual Hochschild complex.

We claim that one can formally derive the complete system of eight coupled
differential equations by variation from the action 
\begin{eqnarray}
S &\sim &\int \{f\circ d_{\bullet }f+\frac 23f\circ f\circ f+g\circ
d_{\otimes }g+\frac 23g\circ g\circ g  \label{18} \\
&&+\varphi \circ d_{\bullet }\varphi +\frac 23\varphi \circ \varphi \circ
\varphi +\psi \circ d_{\otimes }\psi +\frac 23\psi \circ \psi \circ \psi 
\nonumber \\
&&+\lambda _1\circ ^2[d_{\otimes }^{\otimes ^2}f-d_{\bullet }^{\otimes
^2}g+f\otimes f-g\bullet g-\left( f\circ ^2g\right) +\left( g\circ ^2f\right)
\nonumber \\
&&-Comp\left( f,g,g\right) +Comp\left( g,f,f\right) ]  \nonumber \\
&&+\lambda _2\circ ^2[d_{\otimes }^{\otimes ^2}\varphi -d_{\bullet
}^{\otimes ^2}\psi +\varphi \otimes \varphi -\psi \bullet \psi -\left(
\varphi \circ ^2\psi \right) +\left( \psi \circ ^2\varphi \right)  \nonumber
\\
&&-Comp\left( \varphi ,\psi ,\psi \right) +Comp\left( \psi ,\varphi ,\varphi
\right) ]  \nonumber \\
&&+\lambda _3\circ [d_{\bullet }^{\mathcal{F}}\Omega +f\left( \mathcal{F}%
^{\otimes ^2}\right) -\mathcal{F}\left( \varphi \right) +f\circ ^{\mathcal{F}%
}\Omega  \nonumber \\
&&+\Omega \bullet \Omega -\Omega \left( \varphi \right) +f\left( \Omega
^{\otimes ^2}\right) ]  \nonumber \\
&&+\lambda _4\circ [d_{\otimes }^{\mathcal{F}}\Omega +g\left( \mathcal{F}%
^{\otimes ^2}\right) -\mathcal{F}\left( \psi \right) +g\circ ^{\mathcal{F}%
}\Omega  \nonumber \\
&&+\Omega \otimes \Omega -\Omega \left( \psi \right) +g\left( \Omega
^{\otimes ^2}\right) ]\}  \nonumber
\end{eqnarray}
with fields $f,g,\varphi ,\psi ,\Omega ,\lambda _1,\lambda _2,\lambda
_3,\lambda _4$. Here, the fields $\lambda _1,\lambda _2,\lambda _3,\lambda
_4 $ appear as Lagrange multipliers for the constraints where $\lambda
_1,\lambda _2$ are of the same type as $f,g,\varphi ,\psi $, i.e. they are
linear two variable maps on the morphisms of $\mathcal{C}$ and $\mathcal{M}$%
, respectively, while $\lambda _3,\lambda _4$ are of the same type as $%
\Omega $, i.e. one variable linear maps from the morphisms of $\mathcal{M}$
to the morphisms of $\mathcal{C}$.

The action (\ref{18})\ includes interesting special cases: The case 
\[
f=g=\psi =0 
\]
corresponds in the setting of \cite{Bor} to the case where the outer
category $\mathcal{C}$ remains fixed and only the inner ring object $R$ is
deformed. This should correspond to a deformation theory for the field
algebras of \cite{BK}. Imposing an additional commutativity constraint on
the inner ring object $R$ should lead to deformations of usual vertex
algebras (see \cite{Tam} for a study in the setting of the chiral algebras
of \cite{BD}) and - in the infinitesimal case - to the vertex algebra
cohomology of \cite{KV}. It remains a task for future work to rigorously
check the embedding of these cases into our framework.

Since we treated the deformation problem in an abstract setting, one is not
restricted to apply it to the deformations of quantum vertex algebras.
Another special case which is covered by (\ref{18}) is the deformation
problem for three dimensional topological quantum field theories (3d TQFTs
for short). A 3d TQFT is a monoidal functor (\textit{Atiyah functor}) from
the category \textbf{Cobord} of three dimensional cobordisms with disjoint
union as tensor product to the category \textbf{Vect} of finite dimensional
vector spaces with the usual tensor product. The deformation problem for
Atiyah functors is then the deformation problem for this monoidal functor
with \textbf{Cobord} and \textbf{Vect} fixed, i.e it is included into the
deformation theory given by (\ref{18}) by making the special choice $%
\mathcal{M}=\mathbf{Cobord}$ and $\mathcal{C}=\mathbf{Vect}$ and 
\[
f=g=\varphi =\psi =0 
\]
The action (\ref{18}) generalizes the action of the Kodaira-Spencer theory
of gravity of \cite{BCOV} in the following sense: The BV-quantization of
Kodaira-Spencer theory leads to the deformations described by the total
Hochschild complex of the structure sheaf on a complex three dimensional
Calabi-Yau manifold. Physically, this can be seen as the deformation theory
of the BRST-complex of the topological open string. For the topological
closed string a much more complicated deformation theory of the BRST-complex
arises (see \cite{HM}) where the $A_\infty $-deformations - leading to a
Hochschild deformation complex as for the open string case - constitute only
a special case which is very similar to the special case of quantum vertex
algebra deformations where only the inner ring object $R$ is deformed. It is
not clear at present if the action (\ref{18}) leads to the deformations
described in \cite{HM} but it is interesting in any case that the
deformation problem for quantum vertex algebras suggests a generalization of
the action of Kodaira-Spencer theory.

\bigskip

\section{A list of goals}

We now give a list of goals one would like to reach in future work by
studying the action (\ref{18}) in more detail.

\bigskip

\begin{itemize}
\item  Find a gauge fixing for (\ref{18}) similar to the Tian gauge for the
Kodaira-Spencer theory of gravity. Such a gauge fixing is the prerequisite
for making use of (\ref{18})\ in the study of the deformation problem of
quantum vertex algebras.

\item  Study the tree level expansion of (\ref{18}). Can one derive a
formality property for the deformation problem of quantum vertex algebras in
this way as it is possible for Kodaira-Spencer theory on Calabi-Yau
varieties or for the Hochschild complex in the case of \cite{Kon 1997}?

\item  Study the loop corrections for (\ref{18}). Is there an analog of the
holomorphic anomaly of Kodaira-Spencer theory? If yes, to what kind of
structures does the extended moduli space - including non-classical ghost
numbers - lead?\ 

\item  In \cite{Kon 1999} and \cite{KS} it was conjectured that a quotient
of the motivic Galois group which should be given by the
Grothendieck-Teichm\"{u}ller group $GT$, as introduced in \cite{Dri}, acts
as a symmetry on the extended moduli space of Kodaira-Spencer theory. The
algebraic nature of the compatibility constraint appearing between
deformations of $\bullet $ and $\otimes $ suggests that for the deformation
theory of quantum vertex algebras the situation is very similar to the one
studied in \cite{Sch}. We therefore expect that the quantum analogue of $GT$%
, introduced in \cite{Sch} in the form of a self-dual, noncommutative and
noncocommutative Hopf algebra $\mathcal{H}_{GT}$, should appear as a
symmetry on the moduli space related to (\ref{18}). Can one verify this by
studying the field theory given by (\ref{18})?

\item  In \cite{Sch} we have given an argument that deformations on higher
categorical analogs of the structures studied there have to be trivial. We
expect that in a completely similar way the deformation problem of quantum
vertex algebras can not be generalized in a nontrivial way to a similar
deformation problem in the setting of monoidal 2-categories or higher
monoidal $n$-categories. This should imply an unusual stability property for
the field theory given by (\ref{18}) against further deformations. We have
seen that Maurer-Cartan type equations and their generalizations studied in
this paper are strongly related to a perturbative treatment of the
deformation problem (in the sense of deformations given by formal power
series). Would such a stability property mean that the field theory given by
(\ref{18}) can be completely understood from a perturbative treatment? Does
it mean that (\ref{18}) is free of anomalies at any loop order and that the 
\textit{extended} moduli space of (\ref{18}) involves a more or less trivial
extension?

\item Study (\ref{18}) in a concrete example. E.g., study the deformation 
problem for the chiral de Rham complex on the affine space (as introduced 
in \cite{MSV}), i.e. the deformation problem for the tensor product of the 
Heisenberg and the Clifford vertex algebra. This should be of interest for 
studying the deformations of the models appearing in \cite{Wit 2005}.
\end{itemize}

\bigskip

\section{Conclusion}

We have studied the deformation problem of quantum vertex algebras in an
abstract - or better boiled down - setting. We have derived a coupled system
of eight differential equations generalizing the Maurer-Cartan equation of
the usual Hochschild complex. We have shown that this system of equations
can formally be derived from an action principle which in a certain sense
generalizes the action of Kodaira-Spencer theory.

One of the next steps in our work will be the study of (\ref{18}) in a
highly simplified example. For this, we will study the deformation problem 
for the case of topological minimal models where it does reduce to a deformation problem in usual Hochschild cohomology which can be formulated as a deformation problem for a prepotential on a noncommutative moduli space.

\bigskip

\textbf{Acknowledgments:} We would like to thank B. Bakalov, E. Frenkel, M.
Kreuzer, and F. Malikov for discussions on or related to the material of
this paper.


\bigskip

\bigskip

\begin{thebibliography}{Kon 1994}
\bibitem[BCOV]{BCOV}  M. Bershadsky, S. Cecotti, H. Ooguri, C. Vafa, \textit{%
Kodaira-Spencer theory of gravity and exact results for quantum string
amplitudes}, Comm. Math. Phys. 165 (1994), 311-427, hep-th/9309140.

\bibitem[BD]{BD}  A. Beilinson, V. Drinfeld, \textit{Chiral algebras},
American Mathematical Society, Providence 2004.

\bibitem[BK]{BK}  B. Bakalov, V. Kac, \textit{Field algebras},
math.QA/0204282.

\bibitem[Bor]{Bor}  R. E. Borcherds, \textit{Quantum vertex algebras},
math.QA/9903038.

\bibitem[CF]{CF}  L. Crane, I. B. Frenkel, \textit{Four-dimensional
topological quantum field theory, Hopf categories and the canonical bases},
J. Math. Phys. 35, 5136-5154 (1994).

\bibitem[CP]{CP}  V. Chari, A. Pressley, \textit{Quantum groups}, Cambridge
University Press, Cambridge 1994.

\bibitem[Dri]{Dri}  V. G. Drinfeld, \textit{On quasi-triangular Quasi-Hopf
algebras and a group closely related with Gal(}$\overline{\Bbb{Q}}/\Bbb{Q}$%
\textit{)}, Leningrad Math. J., \textbf{2}, 829-860 (1991).

\bibitem[Dub]{Dub}  B. Dubrovin, \textit{Geometry of 2D topological field
theories}, in Springer LNM 1620 (1996), 120-348.

\bibitem[HM]{HM}  C. Hofman, W. K. Ma, \textit{Deformations of closed
strings and topological open membranes}, hep-th/0102201v3.

\bibitem[KO]{KO}  A. Kapustin, D. Orlov, \textit{Vertex algebras, mirrory
symmetry, and D-branes: The case of complex tori}, hep-th/0010293v2.

\bibitem[Kon 1994]{Kon 1994}  M. Kontsevich, \textit{Homological algebra of
mirror symmetry}, alg-geom 9411018.

\bibitem[Kon 1997]{Kon 1997}  M. Kontsevich, \textit{Deformation
quantization of Poisson manifolds I}, math/9709180.

\bibitem[Kon 1999]{Kon 1999}  M. Kontsevich, \textit{Operads and motives in
deformation quantization}, math.QA/9904055.

\bibitem[KS]{KS}  M. Kontsevich, Y. Soibelman, \textit{Deformations of
algebras over operads and Deligne's conjecture}, math.QA/0001151v2.

\bibitem[KV]{KV}  T. Kimura, A. A. Voronov, \textit{The cohomology of
algebras over moduli spaces}, hep-th/9410108v2.

\bibitem[Mac]{Mac}  S. Mac Lane, \textit{Categories for the working
mathematician}, Springer, Berlin 1971.

\bibitem[MSV]{MSV}  F. Malikov, V. Schechtman, A. Vaintrob, \textit{Chiral
de Rham complex}, math.AG/9803041v7.

\bibitem[Sch]{Sch}  K.-G. Schlesinger, \textit{A quantum analogue of the
Grothendieck-Teichm\"{u}ller group}, J. Phys. A: Math. Gen. 35 (2002),
10189-10196, math.QA/0104275.

\bibitem[Tam]{Tam}  D. Tamarkin, \textit{Deformations of chiral algebras},
math.QA/0304211.

\bibitem[Wit 1991]{Wit}  E. Witten, \textit{Mirror manifolds and topological
field theory}, in: S. T. Yau, \textit{Essays on mirror manifolds},
International Press, Hong Kong 1991.

\bibitem[Wit 2005]{Wit 2005}  E. Witten, \textit{Two dimensional models with
(0,2) supersymmetry: Perturbative aspects}, hep-th/0504078v2.
\end{thebibliography}
\end{document}